\documentclass[a4papaer,12pt]{article}
\usepackage{amssymb,amscd,amsmath,amsxtra}
\begin{document}
\voffset = -5cc
\hoffset = -3cc
\setcounter{page}{1}
\raggedbottom
\rm
\parindent 20pt

\ \\

\begin{center}
{\Large{\bf Three-point correlation functions in \\

\ \\

$N=1$ Super Liouville Theory}}\\

\ \\

{\bf R.C.Rashkov}\footnote{
e-mail: rash@@phys.uni-sofia.bg}  \\
\ \\
Department of Theoretical Physics\\
Sofia University, 5 J.Bourchier Blvd\\
1126 Sofia, Bulgaria

\ \\

{\bf M. Stanishkov}\footnote{
e-mail: marian@@bgearn.acad.bg}\\
\ \\
Institute for Nuclear Research\\
and Nuclear Energy\\
Blvd. Tzatigradsko shosse 72\\
1784 Sofia, Bulgaria

\ \\

\end{center}

\ \\

\begin{abstract}
In this letter we propose exact three-point correlation functions
for $N=1$ supersymmetric Liouville theory. Along the lines of \cite{ZZ} we
propose a generalized special function which describe the three-point
amplitudes. We consider briefly the so called reflection amplitudes in the
supersymmetric case.
\end{abstract}

\section{Introduction}
One of the main problems in string theory is how to reduce the theory to
a realistic $D=4$. In his remarcable paper Polyakov \cite{P} showed that the answer
lies in solving the 2D quantum Liouville field theory. Many attempts to
solve this problem have been made in the course of the years, but some
significant success in the subject up to now seems to be absent. There are
several approaches to 2D quantum Liouville theory, namely continuum formulation
(path integral and operator approaches)\cite{DK,P1,Da,Cu}, matrix models
\cite{Br} and
topological field theory \cite{W}. The interest have been oriented
also to conformal matter coupled to Liouville theory \cite{GL}.
It is well known that assuming free
field operator product expansion for the Liouville field the scaling dimensions
found are in complete agreement with the results from the different approaches.
Nevertheless, the Liouville dynamics is not that of a free field.

Recently two important results in the bosonic Liouville theory have been
obtained.
The first one consists in expressing the correlators of matter theory coupled
to gravity as free field ones \cite{AoD,Do}. For this purpose one can expand the
correlation function in powers of the cosmological constant. The integration
over the constant Liouville mode leads to interpretation of the powers of the
Liouville exponential interaction as a screening charge. This allows to
use Dotsenko-Fateev thechnique in Coulomb gas picture. The three-point function
has been computed in different approaches and it has been found to be in
agreement with the results from matrix models \cite{Br} and \cite{AoD,Do}.

The second remarcable result is the exact three-point function of Liouville
vertex operators which has been proposed in \cite{OD} and independently in
\cite{ZZ}.
This result represents an important basis for further investigation of the
conformal blocks and their factorization properties.

In this letter we are extending the results of \cite{OD,ZZ} to the case of $N=1$
supergravity using the super Liouville approach.

One advantage of this approach is that whatever thechnique is
used in the bosonic
case, at least conceptually, can be generalized to the supersymmetric case.
Such computations as far as we know have no analogue in the matrix model
formulation of 2D supergravity or superconformal matter coupled to super
Liouville theory. Moreover the supersymmetric theory  has richer field
content than the bosonic one.

This paper is organized as follows. In Section 1 some basic facts and notations
from super Liouville theory are introduced. In Sections 2 and 3 we propose
exact expressions for the three-point correlation functions in Neveu-Schwarz and
Ramond sectors of the theory. We have generalized the Zamolodchikov's Upsilon
function to our case and some properties of our $\mathcal R(x,a)$ function
are discussed. For some special values of the parameter {\it a} $\mathcal R(x,a)$
coincide with $\Upsilon (x)$.

The pole structure of the three-point function is considered. Using
the reflection properties of the Liouville vertex operators
$V_{\alpha}=V_{Q-\alpha}; R_\alpha=R_{Q-\alpha}$
we introduce the so called reflection
amplitudes in Neveu-Schwarz and Ramond sectors.

\section{Super Liouville theory}
This section is a review of some basic results about super Liouville field
theory. We shall formulate the supersymmetric theory by the action:
\begin{equation}
S_{SL}=\frac {1}{4\pi} \int\hat E\left[\frac 12 D_\alpha
\Phi D^\alpha\Phi - Q\hat R\Phi + \mu e^{b\Phi}\right]
\end{equation}
where the real superfield $\Phi$ possess the expansion:
$$
\Phi = \phi +\theta\psi +\bar\theta\bar\psi+\theta\bar\theta F
$$
Following \cite{DK} we have chosen a background zweibein $\hat E$ ($\hat R$ is the
scalar curvature corresponding to the background metric) and $\mu$ in (1)
is the cosmological constant. The classical equations of motion for (1)
are\footnote{One can choose the background metric to be flat and therefore
$\hat R$ will not be essential in the sequal.}:
\begin{equation}
D_\alpha D^\alpha\Phi = Q\hat R +\mu e^{b\Phi}
\end{equation}
The superspace notations that we shall use are:
\begin{align}
& Z=(z,\theta)\\
& Z_1 -Z_2 = z_1 -z_2 -\theta_1\theta_2
\end{align}
The super energy-momentum tensor of the super Liouville theory is expressed in
terms of the real superfield $\Phi$:
\begin{equation}
T_{SL}=-\frac 12 D\Phi\partial\Phi +\frac Q2 D\partial\Phi
\end{equation}
the central charge of the super-Virasoro algebra being given by:
\begin{equation}
\hat c=1+2Q^2
\end{equation}
The superconformal primary fields are divided in two sector depending
on the boundary conditions of the supercurrent. In the Neveu-Schwarz
sector they are represented by the vertex operators:
\begin{equation}
V_\alpha=e^{\alpha\Phi(Z,\bar Z)}
\end{equation}
of dimension $\Delta_\alpha =\frac 12\alpha(Q-\alpha)$ and in the Ramond
sector by:
\begin{equation}
R_\alpha^\epsilon =\sigma^\epsilon e^{\alpha\Phi(z,\bar z)}
\end{equation}
where $\sigma^\epsilon$ is so called spin field and
$\Delta_\alpha=\frac 12\alpha(Q-\alpha)+\frac{1}{16}$ ($\epsilon =\pm$).
The requirement for the cosmological term in (1) to be (1/2,1/2) form in
order to be able to integrate over the surface, gives a connection between
$Q$ and $b$:
\begin{equation}
Q=b+\frac 1b
\end{equation}
It is easy to see that the operators $V_\alpha=e^{\alpha\Phi}$ and
$V_{Q-\alpha}=e^{(Q-\alpha)\Phi}$ have equal dimensions and therefore they are
reflection image of each other (the same is true also for $R_\alpha^\epsilon$ and
$R_{Q-\alpha}^{\epsilon}$).

As in the bosonic case we can impose the fixed
area condition (which in this case is of dimension of a lenght
rather than of an area). Following
the considerations in \cite{GL} we can impose the above condition
inserting a $\delta$-function into the path integral. The integration over
the constant mode $\phi_0$ from 0 to $\infty$ (the Liouville superfield
$\Phi$ decomposes as follows: $\Phi (Z)=\phi_0 +\Phi ' (Z)$) gives:
\begin{equation}
\langle\prod\limits_{i=1}^{N}e^{\alpha_i\Phi (Z_i)}\rangle =
\left(\frac{\mu}{2\pi}\right)^s\frac{\Gamma (-s)}{b}
\langle\left (\int\hat E e^{b\Phi '}\right )^s\prod\limits_{i=1}^{N}
e^{\alpha_i\Phi ' (Z_i)}\rangle_{S_{SL} '}
\end{equation}
where:
\begin{align}
& S_{SL} '=\frac{1}{4\pi}\int\hat E\left[\frac 12 D_\alpha\Phi 'D^\alpha\Phi '
-Q\hat R\Phi '\right]\\
& sb=Q-\sum\limits_{i=1}^{N}\alpha_i
\end{align}
which reveals the $\mu$-dependence of the correlation function.
We point out that the correlation function on the rhs in (10) is with respect
to the free superfield action. Following the same strategy as in the
bosonic case one has to evaluate the above amplitude as for free field but
$s$ is supposed to be positive integer and then perform an analytical
continuation with respect to $s$.
In this picture the cosmological
term appears as a screening charge. Fortunately, for $N=1$ supersymmetric
conformal theories the super Coulomb gas formalism and supersymmetric
Dotsenko-Fateev integrals are well developed \cite{Kit}.

\section{The exact three-point function in the Neveu-Schwarz sector}
Here we shall follow the approach of [10]which is a slightly modified version
of the one described above.
Consider first the three-point correlation function of Liouville vertex
operators from Neveu-Schwarz sector. The perturbative expansion in the
cosmological constant $\mu$ is given by:

\begin{multline}
\langle V_{\alpha_1}(Z_1)V_{\alpha_2}(Z_2)V_{\alpha_3}(Z_3)\rangle \\
=\int D_{\hat E}\Phi e^{-S_{SL}}e^{\alpha_1\Phi (Z_1)}e^{\alpha_2\Phi (Z_2)}
e^{\alpha_3\Phi (Z_3)}\\
=\sum\limits_{s=0}^\infty\left (\frac{\mu}{2\pi}\right)^s\frac{1}{s!}
\int D_{\hat E} \Phi
e^{-S_{SL} '}\left (\int\hat E e^{b\Phi}\right)^s\prod\limits_{i=1}^{3}
e^{\alpha_i\Phi (Z_i)}
\end{multline}
where the free superfield action $S_{SL} '$ is as in (11).
Specializing to the case of correlation functions on the sphere we
shall concenrate the curvature at infinity ($\infty$,0) and the
considerations will be done for flat zweibein on the plane. Therefore,
we can use the super Coulomb gas formalism in order to evaluate the
correlation function on the rhs in (13). As it is well known, (13) is nonzero
only if:
\begin{equation}
sb=Q-\sum\limits_{i=1}^3\alpha_i
\end{equation}
for any order $s$ of the pertubation series (13).
The result for the $s^{th}$ term in the expansion (13) is [11,12]:
\begin{multline}
\langle V_{\alpha_1}(Z_1)V_{\alpha_2}(Z_2)V_{\alpha_3}(Z_3)\rangle_s \\
=\left (\frac{\mu}{2\pi}\right)^s\frac{1}{s!}
\prod\limits_{i<j}^{3}|Z_i -Z_j|^{-2\alpha_i\alpha_j}\\
\times \int\prod\limits_{j=1}^{s}D^2Y_j\prod\limits_{i=1}^{3}
|Z_i-Y_j|^{-2b\alpha_i}\prod\limits_{i<j}^{s}|Y_i-Y_j|^{-2b^2}
\end{multline}
For $N=1$ case there exists a supersymmetric extension of the
Dotsenko-Fateev integrals \cite{Kit} and an analogous integral expression for the
structure constants can be extracted. Applied to our problem this
integral expression gives for the three-point function in the case of
integer number of screening charges the following result \cite{A-G}:
\begin{align}
& \langle V_{\alpha_1}(Z_1)V_{\alpha_2}(Z_2)V_{\alpha_3}(Z_3)\rangle_s \\ \notag
&=\left (\frac{\mu}{8}
\Delta\left(\frac{b^2}{2}+\frac 12\right)\right)^s
\prod\limits_{i<j}^{3}|Z_i -Z_j|^{-2\delta_{ij}}
\prod\limits_{j=1}^{s}\Delta\left (\frac j2 -\left[\frac j2\right]
-j\frac{b^2}{2}\right)\\ \notag
&\times\prod\limits_{j=0}^{s-1}\prod\limits_{i=1}^{3}
\Delta\left (1-\frac j2 +\left [\frac j2\right]-b\alpha_i -
j\frac{b^2}{2}\right)\times
\begin{cases} 1 \qquad\qquad s\in 2\Bbb N \\
       \frac{\eta\bar\eta}{b^2}\qquad s\in 2\Bbb N+1
\end{cases}
\end{align}
where:
\begin{align}
& \Delta (x)=\frac{\Gamma (x)}{\Gamma (1-x)};\\ \notag
& \delta_{ij}=\Delta_i +\Delta_j -\Delta_k; \quad i\neq j\neq k \\ \notag
& \Delta_i = \alpha_i (Q-\alpha_i) \\ \notag
& \eta =\frac{\theta_1(Z_2-Z_3)+\theta_2(Z_3-Z_1)+\theta_3(Z_1-Z_2)+
\theta_1\theta_2\theta_3}{\left[(Z_1-Z_2)(Z_2-Z_3)(Z_3-Z_1)\right]^{\frac 12}}
\\ \notag
\end{align}
In the above formula we have denoted by $\eta$ the $SL(2|1)$ odd invariant
for any given three
points ($Z_1,Z_2,Z_3$). In contrast to the bosonic case here the correlation
function is different for $s\in 2\Bbb N$ and $s\in 2\Bbb N+1$.

At this point we use the interpretation of the (14) along the lines
of \cite{GL,Po}. It was suggested to consider (14) as a kind of "on-mass-shell"
condition for the exact correlation function. This condition is means that
the exact correlation function sould satisfy the condition:
\begin{equation}
\underset {\sum_{i=1}^{3}\alpha_i=Q-sb}{res}
\langle V_{\alpha_1}V_{\alpha_2}V_{\alpha_3}\rangle =
\frac{\left(-\mu\right)^s}{s!}
\langle V_{\alpha_1}V_{\alpha_2}V_{\alpha_3}
\underbrace{\int\hat E V_b\dots \int\hat E V_b}_s\rangle_{\sum_{i=1}^{3}
\alpha_i=Q-sb}
\end{equation}
when (14) holds for $s=0,1,2\dots$. In general (18) alone seems to be
unsufficient to determine $N$-point function, but for
three Liouville vertex operators the situation is simple: the
coordinate dependence on the left hand side and right hand side is as in the
three-point function (13). Therefore we have the following "on-mass-shell"
condition for the structure constants:
\begin{equation}
\underset {sb=Q-\sum_i\alpha_i}{res}C^{even(odd)}
(\alpha_1,\alpha_2,\alpha_3)=I_s^{even(odd)}(\alpha_1,\alpha_2,\alpha_3)
\end{equation}
where we have denoted by $I_s^{even(odd)}(\alpha_1,\alpha_2,\alpha_3)$
the coordinate independent part of the $s^{th}$ term in the expansion (16).

Now we have to generalize the special function $\Upsilon (x)$ introduced
in \cite{ZZ}. For both, bosonic and supersymmetric cases, we define the
function ($0<Re(x)<Q$):
\begin{align}
 log\mathcal R(x,a)= & \frac 12\int\limits_0^{\infty}\frac{dt}{t}
\left\{\left[\left(\frac Q2-x\right)^2+\left(\frac Q2-a\right)^2
\right]e^{-t}\right.\\ \notag
& \left. -2\frac{sh^2\left[\left(\frac Q2-x\right)+\left(\frac Q2-a\right)
\right]\frac t4+sh^2\left[\left(\frac Q2-x\right)-\left(\frac Q2-a\right)
\right]\frac t4}{sh\frac{t}{2b}sh\frac{bt}{2}}\right\}
\end{align}
The simplest properties that are clear from (20) are:
\begin{align}
& \mathcal R (\frac Q2,\frac Q2)=1 \\ \notag
& \mathcal R (x,a)= \mathcal R (Q-x,a) \\ \notag
& \mathcal R (x,a)= \mathcal R (a,x)
\end{align}
We define also:
\begin{equation}
\mathcal R_0= \left.\frac{d \mathcal R (x,a)}{dx}\right|_{x=a=0} \notag
\end{equation}
We propose the following expression
as an exact three-point function in supersymmetric Liouville theory:
\begin{align}
& C^{even}(\alpha_1,\alpha_2,\alpha_3)= 
\left[\frac{\mu}{8}\Delta\left(\frac{b^2}{2}+\frac 12\right)
b^{-1-b^2}\right]^{\frac{Q-\sum_i\alpha_i}{b}}\\ \notag
& \\ \notag
& \times\frac{\mathcal R_0 \mathcal R(2\alpha_1,0) \mathcal R(2\alpha_2,0)
\mathcal R(2\alpha_3,0)}{\mathcal R(\alpha_1+\alpha_2+\alpha_3 -Q,0)
\mathcal R(x_1,0) \mathcal R(x_2,0) \mathcal R(x_3,0)}
\end{align}

\begin{align}
& C^{odd}(\alpha_1,\alpha_2,\alpha_3)= 
\left[\frac{\mu}{8}\Delta\left(\frac{b^2}{2}+\frac 12\right)
b^{-1-b^2}\right]^{\frac{Q-\sum_i\alpha_i}{b}}\\ \notag
& \\ \notag
& \times\frac{\mathcal R_0 \mathcal R(2\alpha_1,0) \mathcal R(2\alpha_2,0)
\mathcal R(2\alpha_3,0)}{\mathcal R(\alpha_1+\alpha_2+\alpha_3 -Q,b)
\mathcal R(x_1,b) \mathcal R(x_2,b) \mathcal R(x_3,b)}
\end{align}
where:
\begin{equation}
x_i=\alpha_j+\alpha_k-\alpha_i; \qquad i\ne j\ne k
\end{equation}

Thus, we have in general:
\begin{equation}
\langle V_{\alpha_1}V_{\alpha_2}V_{\alpha_3}\rangle =
\left( C^{even}(\alpha_1,\alpha_2,\alpha_3) +
\eta\bar\eta C^{odd}(\alpha_1,\alpha_2,\alpha_3)\right)
\prod\limits_{i<j}|Z_i-Z_j|^{\delta_{ij}}
\end{equation}
This expression for the exact three-point function is
based on the properties of the defined $\mathcal R(x,a)$ function described
below.

We pass to the Ramond sector leaving the reflection amplitudes and the
functional properties of $\mathcal R(x,a)$ to the Section 5.
\section{Three-point function in the Ramond Sector}
According to the explicit form of the Ramond vertex operator (8) its three-point
function has the following perturbative expansion:
\begin{align}
&\langle R^{\epsilon_1}_{\alpha_1}(z_1)
R^{\epsilon_2}_{\alpha_2}(z_2)V_{\alpha_3}
(Z_3)\rangle = \sum\limits_{s=0}^\infty\int du_1\dots du_s\langle
\prod\limits_{i=1}^3 e^{\alpha_i\phi(z_i)}\prod\limits_{j=1}^s e^{b\phi(u_j)}
\rangle \\ \notag
&\times \langle\sigma^{\epsilon_1}(z_1)\sigma^{\epsilon_2}(z_2)
\left (1+\theta_3\bar\theta_3\psi(z_3)\right)
\prod\limits_{j=1}^s\psi(u_j)\rangle
\end{align}
As before, in order the free bosonic correlator to be nonzero we have to impose
the condition (14). It can be interpreted again as a "on-mass-shell" condition
for the exact correlation function. The explicit expression for the integrals in
(26) can be extracted from the corresponding formulae for the Ramond fields
\cite{PZ}. The final result is:
\begin{align}
& \langle R^{\epsilon_1}_{\alpha_1}(Z_1)R^{\epsilon_2}_{\alpha_2}(Z_2)
V_{\alpha_3}(Z_3)\rangle_s
=\left (\frac{\mu}{8}
\Delta\left(\frac{b^2}{2}+\frac 12\right)\right)^s
\prod\limits_{i<j}^{3}|Z_i -Z_j|^{-2\delta_{ij}}\\ \notag
& \times\prod\limits_{j=1}^{s}\Delta\left (\frac j2 -\left[\frac j2\right]
-j\frac{b^2}{2}\right)\prod\limits_{j=0}^{s-1}\prod\limits_{i=1}^{2}
\Delta\left (1+\frac j2 -\left [\frac j2\right]-b\alpha_i -
j\frac{b^2}{2}-\frac 12\right)\\ \notag
& \times\Delta\left (1-\frac j2 +\left [\frac j2\right]-b\alpha_3 -
j\frac{b^2}{2}\right)\times A_{\epsilon_1,\epsilon_2}
\end{align}
where:
\begin{align}
& A_{\epsilon,\epsilon}=
\begin{cases} 1 ,\qquad \qquad\qquad\quad s= 2\Bbb N\\
      \theta_3\bar\theta_3\frac{|z_1-z_3||z_2-z_3|}{|z_1-z_2|},\quad s=2\Bbb N+1
\end{cases}
\end{align}
\begin{align}
A_{\epsilon,-\epsilon}=
\begin{cases} 1 ,\qquad \qquad\qquad\quad s=2\Bbb N+1\\
      \theta_3\bar\theta_3\frac{|z_1-z_3||z_2-z_3|}{|z_1-z_2|};\quad s=2\Bbb N
\end{cases}
\end{align}
and $\Delta(x),\,\, \delta_{ij},\,\,\Delta_i$ as in (17).
Finally, we propose the following expression for the exact three-point function
in the Ramond sector:
\begin{equation}
\langle R^{\epsilon_1}_{\alpha_1}(z_1)R^{\epsilon_2}_{\alpha_2}(z_2)
V_{\alpha_3}(Z_3)\rangle = \left(C^{\epsilon_1,\epsilon_2}+
\theta_3\bar\theta_3\frac{|z_1-z_3||z_2-z_3|}{|z_1-z_2|}
\tilde C^{\epsilon_1,\epsilon_2}\right)\prod\limits_{i<j}|z_i-z_j|^{\delta_{ij}}
\end{equation}
where:
\begin{align}
& C^{\epsilon,\epsilon}(\alpha_1,\alpha_2,\alpha_3)=
\left[\frac{\mu}{8}\Delta\left(\frac{b^2}{2}+\frac 12\right)
b^{-1-b^2}\right]^{\frac{Q-\sum_i\alpha_i}{b}}\\ \notag
& \\ \notag
& \times\frac{\mathcal R_0 \mathcal R(2\alpha_1,b) \mathcal R(2\alpha_2,b)
\mathcal R(2\alpha_3,0)}{\mathcal R(\alpha_1+\alpha_2+\alpha_3 -Q,0)
\mathcal R(x_1,b) \mathcal R(x_2,b) \mathcal R(x_3,0)}
\end{align}
\begin{align}
& C^{\epsilon,-\epsilon}(\alpha_1,\alpha_2,\alpha_3)=
\left[\frac{\mu}{8}\Delta\left(\frac{b^2}{2}+\frac 12\right)
b^{-1-b^2}\right]^{\frac{Q-\sum_i\alpha_i}{b}}\\ \notag
& \\ \notag
& \times\frac{\mathcal R_0 \mathcal R(2\alpha_1,b) \mathcal R(2\alpha_2,b)
\mathcal R(2\alpha_3,0)}{\mathcal R(\alpha_1+\alpha_2+\alpha_3 -Q,b)
\mathcal R(x_1,0) \mathcal R(x_2,0) \mathcal R(x_3,b)}
\end{align}
and $\tilde C^{\epsilon_1,\epsilon_2}$ can be determined using the
supersymmetry ($x_i$ is as in (24)).
\section{Pole structure and reflection amplitudes}
In this section we are going to discuss the pole structure of the
three-point correlation function and to define the so called reflection
amplitudes.
For this purpose we start with some transformation properties
and some functional relations for $\mathcal R(x,a)$ defined in section 2 (see
(20)). Some of the properties are given by (21). Using the integral
representation of $\mathcal R(x,a)$ one can check that the following functional
relation holds:
\begin{equation}
\mathcal R(x+b,a)=b^{-bx+ab}\Delta\left(\frac{bx-ba+1}{2}\right)\mathcal R(x,a+b)
\end{equation}
It is clear that due to the "self-duallity" of $\mathcal R(x,a)$
(i.e. the invariance under $b\to 1/b$) one can conclude that:
\begin{equation}
\mathcal R(x+1/b,a)=b^{\frac xb-\frac ab}\Delta\left(\frac{x-a+b}{2b}\right)
\mathcal R(x,a+1/b)
\end{equation}
In the expressions for the correlation functions
(22,23,31,32) the following special cases of $\mathcal R(x,a)$ are used:
\begin{align}
& \mathcal R(x,0)=\mathcal R(x,Q)=\Upsilon_1 (x)\\
& \mathcal R(x,b)=\mathcal R(x,Q-b)=\Upsilon_2 (x)
\end{align}
Using the above functional relations it is easy to prove that:
\begin{align}
& \Upsilon_1(x+b)=b^{-bx}\Delta\left(\frac{bx+1}{2}\right)\Upsilon_2(x)\\
& \Upsilon_2(x+b)=b^{1-bx}\Delta\left(\frac{bx}{2}\right)\Upsilon_1(x)
\end{align}
and
\begin{align}
& \Upsilon_1(x+1/b)=b^{\frac xb}\Delta\left(\frac{x+b}{2b}\right)\Upsilon_2(x)
\\
& \Upsilon_2(x+1/b)=b^{\frac xb-1}\Delta\left(\frac{x}{2b}\right)\Upsilon_1(x)
\end{align}
We would like to point out that:
\begin{equation}
\mathcal R(x,x)=\Upsilon (x)
\end{equation}
where $\Upsilon (x)$ is the Zamolodchikov's Upsilon function \cite{ZZ}.

It is easy to verify that, using the above properties,
$\Upsilon_1(x)=\mathcal R(x,0)$ and
$\Upsilon_2(x)=\mathcal R(x,b)$ are entire functions of $x$ with the following
zero-structure:
\begin{align}
&\Upsilon_1(x)=0\quad for\quad x=-nb-\frac mb;\quad n-m=even\\ \notag
&\Upsilon_2(x)=0\quad for\quad x=-nb-\frac mb;\quad n-m=odd
\end{align}
and due to (21):
\begin{align}
&\Upsilon_1(x)=0\quad for\quad x=(n+1)b+\frac{m+1}{b};\quad n-m=even\\ \notag
&\Upsilon_2(x)=0\quad for\quad x=(n+1)b+\frac{m+1}{b};\quad n-m=odd
\end{align}
($n,m$ are non-negative integers).

Using all the above properties of $\mathcal R(x,a)$ it is straightforward to
check that the proposed exact three-point functions satisfy the
"on-mass-shell" condition (14).

As in the bosonic case \cite{ZZ}, the proposed correlators as a function of
$\alpha=\sum_i^3\alpha_i$
have more poles than expected: at $\alpha=Q-n/b -mb$ and at
$\alpha=2Q+n/b +mb$. They appear when more general multiple integrals
are considered:
\begin{equation}
\underset{\sum_i\alpha_i=Q-\frac nb -mb}{res}
\langle V_{\alpha_1}V_{\alpha_2}V_{\alpha_3}\rangle=
\frac{(\tilde\mu)^n(\mu)^m}{n!m!}
\langle\prod\limits_{i=1}^{3}V_{\alpha_i}(Z_i)\prod\limits_{k=1}^{n}\int
V_{1/b}(X_k)\prod\limits_{l=1}^{m}\int V_{b}(Y_l)\rangle
\end{equation}
where:
\begin{equation}
\frac{\tilde\mu}{8}\Delta\left(\frac{1}{2b^2}+\frac 12\right)=
\left(\frac{\mu}{8}\Delta\left(\frac{b^2}{2}+\frac
12\right)\right)^{\frac{1}{b^2}}
\end{equation}
We note that the correlation function (44) is self-dual with respect
to $b\to\frac 1b, \mu\to\tilde\mu$.
As it was mentioned in Section 1 the Liouville vertex operators $V_\alpha$
and $V_{Q-\alpha}$ are reflection image of each other. We shall use this
property in order to define the so called reflection amplitudes in the
supersymmetric case.

We start with the reflection amplitude in the Neveu-Schwarz sector. Due to
the reflection properties we define:
\begin{equation}
C^{even(odd)}(Q-\alpha_1,\alpha_2,\alpha_3)=
S^{NS}(\alpha_1)C^{even(odd)}(\alpha_1,\alpha_2,\alpha_3)
\end{equation}
where $S^{NS}(\alpha_1)$ is the reflection amplitude. In more details,
using (22) we have:
\begin{align}
& C^{even}(Q-\alpha_1,\alpha_2,\alpha_3)=
\left[\frac{\mu}{8}
\Delta\left(\frac{b^2}{2}+\frac 12\right)
b^{-1-b^2}\right]^{\frac{Q-\sum_i\alpha_i}{b}}
\left[\frac{\mu}{8}
\Delta\left(\frac{b^2}{2}+\frac 12\right)
\right]^{\frac{2\alpha_1-Q}{b}}\\ \notag
& \\ \notag
&\times b^{(Q-2\alpha_1)Q} \frac{\mathcal R_0 \mathcal R(2Q-2\alpha_1,0)}
{\mathcal R(Q\underbrace{-\alpha_1+\alpha_2+\alpha_3}_{x_1} -Q,0)
\mathcal R(\alpha_2+\alpha_3+\alpha_1-Q,0)}\\ \notag
& \\ \notag
&\times\frac{ \mathcal R(2\alpha_2,0)\mathcal R(2\alpha_3,0)}
{\mathcal R(Q\underbrace{-\alpha_1+\alpha_3-\alpha_2}_{-x_3},0)
\mathcal R(Q\underbrace{-\alpha_1+\alpha_2-\alpha_3}_{-x_2},0)x_3,0)}\\ \notag
&=
\left[\frac{\mu}{8}
\Delta\left(\frac{b^2}{2}+\frac 12\right)
b^{-1-b^2}\right]^{\frac{Q-\sum_i\alpha_i}{b}}
b^{(Q-2\alpha_1)Q}
\left[\frac{\mu}{8}
\Delta\left(\frac{b^2}{2}+\frac 12\right)
\right]^{\frac{2\alpha_1-Q}{b}}\\ \notag
& \\ \notag
&\times\frac{\mathcal R_0  \mathcal R(2\alpha_2,0)
\mathcal R(2\alpha_3,0)\mathcal R(Q-(2\alpha_1,0-Q))}
{\mathcal R(x_1,0)
\mathcal R(\alpha_2+\alpha_3+\alpha_1-Q,0)\mathcal R(Q-x_3,0)
\mathcal R(Q-x_2,0)}
\end{align}
Due to (21)
\begin{align}
&C^{even}(Q-\alpha_1,\alpha_2,\alpha_3)=
\left[\frac{\mu}{8}
\Delta\left(\frac{b^2}{2}+\frac 12\right)
b^{-1-b^2}\right]^{\frac{Q-\sum_i\alpha_i}{b}}
\left[\frac{\mu}{8}\Delta\left(\frac{b^2}{2}+\frac 12\right)
\right]^{\frac{2\alpha_1-Q}{b}}\\ \notag
& \\ \notag
&\times b^{(Q-2\alpha_1)Q}\frac{\mathcal R_0  \mathcal R(2\alpha_2,0)
\mathcal R(2\alpha_3,0)\mathcal R(2\alpha_1-Q,0)}
{\mathcal R(\alpha_1+\alpha_2+\alpha_3-Q,0)
\mathcal R(x_1,0)
\mathcal R(x_2,0)
\mathcal R(x_3,0)}
\end{align}
Therefore the reflection amplitude is deduced form the above as:
\begin{equation}
b^{(Q-2\alpha_1)Q}
\left[\frac{\mu}{8}
\Delta\left(\frac{b^2}{2}+\frac 12\right)
\right]^{\frac{2\alpha_1-Q}{b}}
\mathcal R(2\alpha_1-Q)=S^{NS}(\alpha_1)\mathcal R(2\alpha_1)
\end{equation}
Repeating the same considerations for $C^{odd}(\alpha_1,\alpha_2,\alpha_3)$
we found the same expression. Using the functional relations of $\mathcal R(x,a)$
we find that the reflection amplitude equals to:
\begin{equation}
S^{NS}(\alpha)=
\left[\frac{\mu}{8}
\Delta\left(\frac{b^2}{2}+\frac 12\right)
\right]^{\frac{2\alpha_1-Q}{b}}
b^{2+\frac{2(Q-2\alpha)}{b}}
\frac{\Delta\left(b\alpha -\frac{b^2}{2}+\frac 12\right)}
{\Delta\left(2-\frac{\alpha}{b}+\frac{1}{2b^2}-\frac 12\right)}
\end{equation}

As in the bosonic case we associate the reflection amplitude with the
two-point correlation function \cite{OD,ZZ}.

Now we pass to the Ramond sector and consider the correlation functions
(31,32) together. In this functions we have two Ramond fields
($R^{\epsilon_1}_{\alpha_1},
R^{\epsilon_2}_{\alpha_2}$) and one Neveu-Schwarz field
($V_{\alpha_3}$). Therefore the
reflection of the first two fields will give us one reflection amplitude, but
the reflection of the third field will differs from the first one.
For instance:
\begin{equation}
C^{\epsilon,\pm\epsilon}(Q-\alpha_1,\alpha_2,\alpha_3)=
S^R(\alpha_1)C^{\epsilon,\mp\epsilon}(\alpha_1,\alpha_2,\alpha_3)
\end{equation}
where:
\begin{equation}
S^R(\alpha_1)=
\left[\frac{\mu}{8}
\Delta\left(\frac{b^2}{2}+\frac 12\right)
\right]^{\frac{2\alpha_1-Q}{b}}
b^{\frac{2(Q-2\alpha)}{b}}
\frac{\Delta\left(b\alpha_1-\frac{b^2}{2}\right)}
{\Delta\left(1-\frac{\alpha_1}{b}+\frac{1}{2b^2}\right)}
\end{equation}
For the reflection of the Neveu-Schwarz field we found:
\begin{equation}
C^{\epsilon,\pm\epsilon}(\alpha_1,\alpha_2,Q-\alpha_3)=
S^{NS}(\alpha_3)C^{\epsilon\pm\epsilon}(\alpha_1,\alpha_2,\alpha_3)
\end{equation}
where $S^{NS}(\alpha)$ is as in (50).

Once again we would like to mention that, as in the bosonic case,
the reflection amplitudes are associated with the two-point correlation
function. The two reflection amplitudes found in (52,53) correspond to the
correlator of two Ramond fields and of two Neveu-Schwarz fields respectively.
This can be seen if we consider the correlation function of two fields
devided by the volume of the group leaving invariant two marked points.
Using the mechanism of canceled integration in the path integral formulation
, as it has been done in \cite{OD}, we can find an expression in terms of
$\mathcal R(x,a)$.
After some simple but tedious transormations one can arrive to the expressions
(50,52). We leave the detailed calculations to a forthcomming paper \cite{RS}.

\ \\

{\bf{\large Conclusion}}

\ \\

We proposed an expression for the exact three-point correlation functions
(22,23,31,32) in the case of super Liouville theory. The corresponding
correlators of the Neveu-Schwarz and Ramond sectors are considered and some
basic properties are discussed. In analogy with the bosonic case the
corresponding reflection amplitudes are introduced. Taking into account that
the proposed correlators (22,23,31,32) are only conjectures, some analysis on
the subject and additional check is needed.
We address this discussion as well as
some detailed calculations and open questions to a forthcomming paper \cite{RS}.

\ \\
{\bf Acknowledgments}
\ \\
Research of R.R. is supported in part by Grant MM402/94 \# Bulgarian Ministry
of Education and Science.

\ \\

\end{document}